\documentclass[aps,prb,notitlepage,twocolumn,superscriptaddress]{revtex4-2}
\usepackage{amsmath}
\usepackage{amssymb}
\usepackage{bbold}
\usepackage{color}
\usepackage{tikz}
\usepackage{pgfplots}
\pgfplotsset{compat=1.5}
\usepackage{graphicx}
\usepackage[colorlinks=true, linkcolor=blue, citecolor=blue, pdfencoding=auto]{hyperref}
\usepackage{blindtext}
\usepackage{physics}

\newcommand{\intBZ}{ \int_{\rm BZ}  }

\begin{document}
\title{
Long and short time linear response of metals: a geometric approach
}
\author{Nishchhal \surname{Verma} }
\affiliation{Department of Physics, Columbia University, New York, NY 10027, USA}

\author{Raquel \surname{Queiroz} }
\email{raquel.queiroz@columbia.edu}
\affiliation{Department of Physics, Columbia University, New York, NY 10027, USA}
\affiliation{Center for Computational Quantum Physics, Flatiron Institute, New York, New York 10010, USA}

\begin{abstract}
The time-dependent quantum geometric tensor, which captures dipole fluctuations of bound electrons, is essential for understanding the electronic properties of insulators, superconductors, and flat bands.
It is often considered subleading for low-energy descriptions of metals that are dominated by intra-band processes. Here, we revisit this perspective and highlight scenarios where the quantum geometry of the wavefunctions close to the Fermi surface plays a significant role.
We compute the time-dependent quantum geometric tensor for metals, explain its divergence, and contrast it against singular geometric tensors of Dirac and Weyl semi-metals. We identify the ratio of Drude to total spectral weight, $D/\mathcal{S}_1$, as a lattice-scale probe of bound versus itinerant charge, and quantify it in the kagome metal, where the two van Hove fillings respond differently despite identical Fermi surfaces.
\end{abstract}

\maketitle

\section{Introduction}
In conventional descriptions of metals, low-energy response is often dominated by quasiparticle dispersion, while geometric properties of electronic wavefunctions play a secondary role. In simple single-band descriptions, low-frequency transport is dominated by intraband processes, and quantities such as the Drude weight are primarily determined by band dispersion rather than explicit wavefunction geometry \cite{ashcroft1977solid}.
Yet metals also contain electronic degrees of freedom that remain locally polarized on short timescales, producing dipole fluctuations governed by the geometry of occupied states.

The widespread applicability of low-energy theories follows from a separation of scales between low- and high-energy electronic states. This separation is most evident in metals where electrons behave as bound at short timescales and itinerant at long times \cite{ashcroft1977solid}. The Landau Fermi liquid theory formalizes this aspect and reveals a universality in the low-energy behavior that exists despite variations in chemical composition and lattice structure of various metals \cite{pines2018theory}.
In contrast, the separation is non-existent in topological systems like Landau levels, where features of the wavefunctions at the high-energy UV-scale lead to rich phenomenology in the low-energy IR-scale \cite{Girvin1987}.
Many quantum materials occupy an intermediate regime where low-energy electronic structure is neither fully captured by a simple separation of scales nor protected by topology.
Nevertheless, their low-energy states can retain nontrivial quantum geometry encoded in wavefunction overlaps and matrix elements.

Broadly speaking, quantum geometry is a formalism to describe changes in the wavefunction as the Hamiltonian is varied parametrically \cite{berry1984quantal, Provost1980}.
In terms of quantum materials, it refers to the variation of wavefunctions in the Brillouin zone which is itself tied to the fluctuation of the position operator in the ground state \cite{Verma2026Review, Yu2025QGmaterials}.
Quantum geometry therefore enters linear-response theory through generalized optical conductivity sum rules that relate integrated response functions to geometric properties of electronic states \cite{Souza2000, onishi2024quantum, Souza2025bounds}.
A connection between conductivity and quantum geometry was recently established through the time-dependent quantum geometric tensor (tQGT) \cite{verma2024instantaneous}, motivated by the relationship between optical response functions, interband dipole matrix elements, and geometric properties of evolving quantum states \cite{Resta2006, Komissarov2024capacitance}.
Various generalizations of quantum geometric response have been explored, including thermoelectric transport \cite{Lhachemi2026} and mixed-state many-body quantum geometry formulated using metrics such as the Bures metric for evolving density matrices \cite{Guan2026, Ji2025}.
These sum rules provide experimentally accessible signatures of phenomena such as metal-insulator transitions \cite{Yan2026ManyBodyQuantumGeometric} and time-reversal symmetry breaking through optical Hall responses \cite{Abelev2026ScalingRelationOpticalHall}.

The real part of $t=0$ tQGT, also known as quantum weight \cite{onishi2024quantum, Onishi2024PRL}, is identical to the quantum Fisher information associated with the position operator at zero temperature \cite{balut2024quantum, Mao2024QS}. Formally, it quantifies the dipole matrix elements between two partitions of the Hilbert space defined by the projector $\hat{P}$ and the complementary projector $\hat{Q}= 1 - \hat{P}$.
This reveals that the quantum weight depends on virtual transitions outside the low-energy subspace and therefore cannot, in general, be determined solely from low-energy bands.
Adding bands above the Fermi level, despite not affecting the ground state, can still change the quantum weight.
This is not unexpected from a sum-rule viewpoint, since adding bands modifies dipole matrix elements and shifts absorption to different frequencies, which in turn affects the sum rule. Because of this, defining an observable for the quantum weight is tricky \cite{Verma2024Step, tam2024corner, onishi2024structure}, though some progress has been made using approximations that include parts of the full Hilbert space \cite{Kang2025}.

The QGT obtained from the sum rule diverges in metals due to the presence of a Fermi surface, the same divergence known from the electronic localization length of the metallic state \cite{Resta1999, Souza2000}. It has also been used to diagnose gapless phases through the static structure factor \cite{Jeon2026, Tam2026FSgeometry}. 
In contrast, the tQGT is not restricted to this divergence. Even though the $t=0$ component is singular, the linear-in-time term defines the charge stiffness, which quantifies the ability of the system to sustain dissipationless charge current \cite{Resta2018Drude}. This quantity remains finite only when low-energy charged excitations exist, and is a defining property of metals \cite{Kohn1964}.

Another feature of metals is the non-commutativity of the limits $q \rightarrow 0$ and $\omega \rightarrow 0$ in various correlation functions.
Discerning between the two order of limits lies beyond the scope of tQGT, which is only defined in the transport limit. While this is not an issue for insulators, where the two limits commute, it leads to discrepancies in metals, among them, a missing term in the Streda response that arises from orbital magnetic moment of the Fermi surface \cite{Haldane2004}.

Motivated by these questions, this work aims to highlight the role of quantum geometry and tQGT in gapless phases. The paper is organized as follows. We begin with a review of response theory, introducing the tQGT and its connection to the continuity equation.
Next, using optical conductivity, we quantify the separation between itinerant and bound electrons, identifying the conditions under which quantum geometry can be considered negligible.
The framework extends to topological Dirac and Weyl semimetals, which lack a clear separation of scales.

\begin{figure}
    \centering
    \includegraphics[width=\columnwidth]{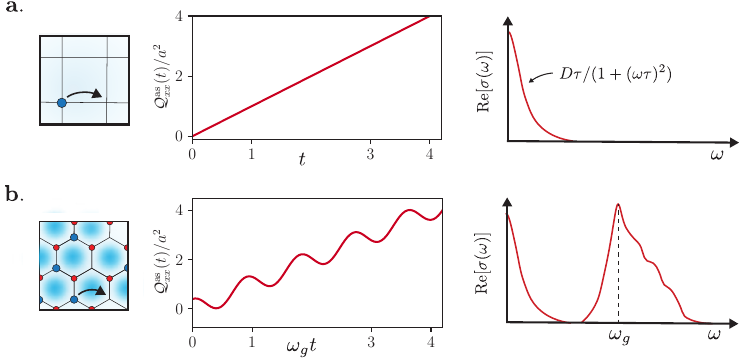}
    \caption{ Itinerant and bound charge in time and frequency. (a) In a single-orbital metal, dipole fluctuations grow linearly in time, $\mathcal{Q}^{\rm as}_{xx}(t) = D\,t$, with the charge stiffness $D$ as the growth rate (middle). The optical conductivity is a single Drude peak, broadened by a finite scattering time $\tau$ into the Lorentzian $D\tau/(1+(\omega\tau)^2)$ (right). (b) In a multi-orbital metal, bound charge adds oscillations at the inter-band scale $\omega_g$ on top of the same linear growth. In frequency, these appear as an inter-band absorption peak centered at $\omega_g$, well separated from the Drude peak. The ratio $D/\mathcal{S}_1$ of Drude to total spectral weight quantifies the fraction of electrons that are itinerant rather than bound.}
    \label{fig:mainIdea}
\end{figure}

\section{Response theory and the tQGT}
We review the density $\rho$, current $J$ and position $\hat{r}$ operators in a generic many-body system described by a Hamiltonian $\mathcal{H}$. The density and the current operators satisfy the continuity equation
\begin{equation}
    \partial_t \rho + \nabla \cdot J = 0 \label{eq:ctyEq:rspace}
\end{equation}
whereas the current and position operators are related by the Heisenberg equation of motion
\begin{equation}
    J = \dfrac{ie}{\hbar} [ \mathcal{H}, \hat{r}].\label{eq:defJ:rspace:comm}
\end{equation}
These relations help compute response functions. Density and current operators typically couple to external probes, while the position operator is related to the quantum geometric tensor.

Of particular interest to our work is the charge response. Under the application of an external electromagnetic field described by a vector potential $A(r,t)$, the current is given by $J = -\delta \mathcal{H}/\delta A$. We have chosen to work in the velocity gauge where both electric and magnetic fields are described by a spatially varying time dependent vector potential $A(r,t)$.
Since the vector potential is gauge-dependent, this choice comes with the necessity of enforcing sum rules.

Let us first consider current-current response. As is well known, different order of limits $ q \to 0 $ and $\omega \to 0$ of the response function can distinguish insulators from metals. Specifically, the response function
\begin{equation}
    \Lambda_{\mu\nu}(q,\omega) = \dfrac{i}{V} \int_0^\infty dt \, e^{i\omega t} \, \left\langle [ j_\mu(q,t), j_\nu(-q,0) ] \right\rangle
\end{equation}
is the coefficient between induced current and applied vector potential
\begin{equation}
    \langle j_\mu(q, \omega) \rangle = ( \langle j_{\mu\nu,D} \rangle - \Lambda_{\mu\nu}(q,\omega) ) A_\nu(q, \omega) \label{eq:curr-A-eq}
\end{equation}
where $ j_{\mu\nu,D} = \delta^2 H/\delta A_{\mu}A_\nu$ is the diamagnetic part of the current operator. The limits of  $ q \to 0 $ and $\omega \to 0$ now correspond to induced current from static or dynamic electro-magnetic fields \cite{Scalapino1993}.

\subsection{Longitudinal Response}
In the presence of a uniform time-varying Electric field, the vector potential can be chosen to be only a function of time. It sets $q = 0$ and $A_\nu(\omega) = E_\nu(\omega)/i\omega$ with the corresponding response
\begin{equation}
    \langle j_\mu(\omega) \rangle = \left(\dfrac{1}{i\omega}\left( \langle j_{\mu\nu,D} \rangle - \Lambda_{\mu\nu}(q=0,\omega) \right)\right) E_\nu( \omega)
\end{equation}
from which one can identify the conductivity $\sigma_{\mu\nu}(\omega)$ as the quantity inside the brackets.
The zero frequency pole of conductivity defines the Drude weight
\begin{equation}
    D_{\mu\nu} = \lim\limits_{\omega \rightarrow 0} \left( \langle j_{\mu\nu,D} \rangle - \Lambda_{\mu\nu}(q=0,\omega) \right)
\end{equation}
which dates back to Drude theory in the limit of infinite scattering time $\tau$. However, this pole is not stable against disorder and broadens immediately to a Lorentzian with width set by the scattering time. Formally, it is referred to as charge stiffness, a name that originates from an equivalent definition in terms of the exact many-body states \cite{Kohn1964}. At zero temperature, the longitudinal part, $D = \sum_\mu D_{\mu\mu}$ can be written as $D = \partial^2 E_0(\phi)/\partial \phi^2$, where $ E_0 $ is the exact many-body ground state energy and $\phi$ is an infinitesimal flux threaded through the system. This definition is equivalent to the one in response theory as a consequence of the Feynman--Hellmann theorem
\begin{equation}
    D = \partial_\phi^2 ( \langle 0 | H | 0 \rangle ) = \langle 0 | \partial_\phi^2 H | 0 \rangle - 2\sum_{m \neq 0} \frac{|\langle m | \partial_\phi H | 0 \rangle|^2}{E_m - E_0}.
\end{equation}
Charge stiffness captures the energy cost of creating an infinitesimal charged excitation.
Although this expression is limited to numerical algorithms that use exact diagonalization, charge stiffness is a physical observable that appears in plasmonic responses.

We now consider a pure gauge field, $A(r,t)$ with $\nabla\cdot A \neq 0$ but $\nabla \times A = 0$ and $\partial_t A = 0$. The formalism, as it stands in Eq.~\eqref{eq:curr-A-eq}, will lead to a fictitious current that must be set to zero. This requirement imposes the condition
\begin{align}
    \langle j_{\mu\nu,D} \rangle &= \Lambda_{\mu\nu}(q\rightarrow 0,\omega=0) \\
    &= \dfrac{1}{\pi}\int\limits_{-\infty}^\infty d\omega \; \dfrac{ {\rm Im}[\Lambda_{\mu\nu}(q\rightarrow 0, \omega)]  }{\omega}
\end{align}
where we used Kramers-Kronig relations to relate the real part of $\Lambda_{\mu\nu}$ to a sum rule of its imaginary part. This relation corresponds to the $f$-sum rule, which must be satisfied by all physical systems as it is based on gauge invariance \cite{Scalapino1993}. An immediate consequence of this sum rule is that the charge stiffness can be written as
\begin{equation}
    D_{\mu\nu} = \lim\limits_{q \rightarrow 0} \lim\limits_{\omega \rightarrow 0} \Lambda_{\mu\nu}(q,\omega)  - \lim\limits_{\omega \rightarrow 0} \lim\limits_{q \rightarrow 0} \Lambda_{\mu\nu}(q,\omega) \label{eq:many:body:drude}
\end{equation}
which emphasizes the order of limits in phases with finite charge stiffness.
This feature is built into diagrammatic perturbation theory where only a Fermi surface can lead to different results with different order of limits. However, we note that this derivation is completely general and does not rely on any assumptions about single-particle theory.
Two comments are in order. First, Eq.~\eqref{eq:many:body:drude} defines the ideal charge stiffness of the clean system; with disorder the pole broadens and the finite Drude weight measured in optics is a distinct quantity. Second, while the order-of-limits statement is fully many-body, the explicit evaluations of Sec.~\ref{sec:fs} assume a Slater-determinant ground state.

\subsection{Time-dependent Quantum Geometric Tensor}
The current operator is the time derivative of the position operator, Eq.~\eqref{eq:defJ:rspace:comm}, which suggests that the conductivity, being a current-current correlation, can be written entirely in terms of position operators. This is achieved by the time dependent quantum geometric tensor (tQGT)
\begin{equation}
    \mathcal{Q}_{\mu\nu}(t) = \left\langle \hat{r}_\mu(t) \;(1-\hat{P})\; \hat{r}_\nu \right\rangle
\end{equation}
where $\hat{P}$ is the projector onto the ground state. At zero temperature and $t=0$, it reduces to the quantum geometric tensor whose real part is the quantum weight and imaginary part is the Chern number.
There are many peculiarities with this correlation function that make it difficult to relate it directly to a response function. We need to anti-symmetrize with time, $\mathcal{Q}^{\rm as}_{\mu\nu}(t) = (\mathcal{Q}_{\mu\nu}(t) - \mathcal{Q}^\dag_{\mu\nu}(t))/i$, to eventually get
\begin{equation}
    \sigma_{\mu\nu}(t) = \dfrac{\pi e^2}{\hbar} \Theta(t) \partial_t \mathcal{Q}^{\rm as}_{\mu\nu}(t). \label{eq:tQGT:cond}
\end{equation}
We next invoke the appearance of charge stiffness in $\sigma_{\mu\nu}(t) = D_{\mu\nu} \Theta(t) + \cdots$ to propose that, in addition to other terms, tQGT must have a linear in time component $\mathcal{Q}_{\mu\nu}(t) = D_{\mu\nu} t$ that quantifies the rate of propagation of dipole fluctuations in time.
A clean metal sustains a ballistic current, $\sigma(t) = D\,\Theta(t)$, and the dipole correlator grows with slope $D$ \cite{verma2024instantaneous}. 

We take a moment to discuss the difference between tQGT and its anti-symmetric component, $\mathcal{Q}^{\rm as}_{\mu\nu}$. It is known that the real part, $\delta_{\mu\nu} \mathcal{Q}_{\mu\nu}(0)$, diverges in metals due to the presence of a Fermi surface \cite{Resta2011TheTheory}.
However, this divergence is absent in both the anti-symmetric part of tQGT and the conductivity.
Interestingly, the anti-symmetric part avoids the divergence while still capturing the charge stiffness.
This structure inherently prevents a straightforward linear-response observable for the quantum weight, necessitating the search for responses that couple to its symmetric part, $\mathcal{Q}^{\rm s}_{\mu\nu}(t) = \mathcal{Q}_{\mu\nu}(t) + \mathcal{Q}^\dag_{\mu\nu}(t)$.
The lack of such an observable for quantum weight is also tied to its connection to the quantum Fisher information \cite{balut2024quantum}.

The tQGT is defined in the transport limit, where $q \rightarrow 0$ is taken first while $\omega$ is kept finite. For the longitudinal response nothing is lost in this restriction, since the opposite order of limits is fixed by the $f$-sum rule. 
Whether the same holds in other channels requires examining responses where the two limits do not commute, which brings us to the Hall response.

\subsection{Hall Response}
We consider the Hall response where a field applied along $x$ produces a current in $y$.
Unlike the longitudinal response, the Hall response $\epsilon_{\mu\nu}\sigma_{\mu\nu}(\omega\rightarrow 0)$ does not have a pole at zero frequency. This is because the diamagnetic current cancels with anti-symmetrization and the remaining part leads to the TKNN formula \cite{TKNN1982} for insulators
\begin{equation}
    \epsilon_{\mu\nu} \dfrac{ {\rm Im}[ \Lambda_{\mu\nu}(q=0,\omega)] }{\omega} = \dfrac{e^2}{h } \sum\limits_{n\in {\rm filled}} \mathcal{C}_n
\end{equation}
where $\mathcal{C}$ is the Chern number, a topological invariant. Since the Chern number originates from the non-commutativity of projected position operators, the insulating Hall response fits naturally within the position-operator representation of the tQGT.

Given the appearance of Fermi surface contributions to the longitudinal conductivity, it is tempting to assert the same for Hall conductivity.
However, there are subtle issues with the order of limits $ \omega,q \rightarrow 0 $. Unlike the longitudinal case, where $\omega\rightarrow0$ and then $q\rightarrow 0$ yielded a sum rule, the Hall response is finite and given by
\begin{equation}
     \lim_{q \to 0} \lim_{\omega \to 0} \left( \epsilon_{\mu\nu} \dfrac{ \Lambda_{\mu\nu}(q,\omega) }{i\omega} \right)
\end{equation}
where $ \epsilon_{\mu\nu} $ is the Levi-Civita tensor.
We compute this expression by first identifying $\Lambda_{\mu\nu}(q,\omega)/i\omega$ with the density-current response function
\begin{equation}
    \chi_{0,\nu}(q,\omega) = \dfrac{i}{V} \int_0^\infty dt \, e^{i\omega t} \, \left\langle [ \rho(q,t), j_\nu(-q,0) ] \right\rangle.
\end{equation}
Using the continuity equation, $\omega \rho(q, \omega) = - q_\mu j_\mu(q, \omega)$, we see that $i\omega \chi_{0,\nu}(q,\omega) = q_\mu \Lambda_{\mu\nu}(q,\omega)$. This identity constrains only the longitudinal projection of $\Lambda_{\mu\nu}$; differentiating it with respect to $q_\mu$ gives $\Lambda_{\mu\nu} + q_\alpha \partial_{q_\mu} \Lambda_{\alpha\nu} = i\omega\, \partial_{q_\mu} \chi_{0,\nu}$, and taking $\omega \rightarrow 0$ first drops the second term. This leads to
\begin{equation}
    \lim_{q \to 0} \lim_{\omega \to 0} \left( \epsilon_{\mu\nu} \dfrac{ \Lambda_{\mu\nu}(q,\omega) }{i\omega} \right) = \epsilon_{\mu\nu} \partial_\mu \chi_{0, \nu}(q)\big|_{q\rightarrow 0},
\end{equation}
which is exactly the Streda formula \cite{Smrcka1977, Streda1982} for Hall conductivity where $ \sigma_H $ is viewed as the linear response of the density due to an applied magnetic field.

Similar to the longitudinal response, we conjecture that the difference between two limits should identify the metallic contribution to the Hall conductivity
\begin{equation}
    \sigma_{H}^{\rm M} =  \epsilon_{\mu\nu} \left(\lim_{q \to 0} \partial_\mu \chi_{0, \nu}(q) - \lim_{\omega \to 0} \dfrac{ \Lambda_{\mu\nu}(\omega) }{i\omega}\right) \label{eq:sigmaH:metal}
\end{equation}
Unlike the longitudinal case, this difference is not protected by a conservation law. The $f$-sum rule pins one order of limits of the longitudinal response through gauge invariance, which is what makes Eq.~\eqref{eq:many:body:drude} exact. In the Hall channel the diamagnetic term drops under anti-symmetrization, no Ward identity constrains the response, and the discarded term $q_\alpha \partial_{q_\mu}\Lambda_{\alpha\nu}$ above is precisely $\sigma_H^{\rm M}$. Eq.~\eqref{eq:sigmaH:metal} is therefore established only at the level of perturbation theory (Appendix~\ref{app:hall}) \cite{Haldane2004, Nagaosa2010}.
This quantity would be finite for metals that break time-reversal and for chiral states that exist at boundaries of topological insulators.

The same structure explains why the metallic Hall response evades the tQGT. The first term in Eq.~\eqref{eq:sigmaH:metal} involves the $q$-derivative of a density response, while the position operator appears through $\rho_q = e^{iq\cdot \hat{r}}$, making the connection unclear as the derivative acts outside the response function.
This issue does not arise in the second term, where position operators explicitly appear and evaluate to the Chern number. In sum, the Hall response can be written in terms of position operators only when the two limits commute, that is, when the metallic correction $\sigma_H^{\rm M}$ vanishes.

To conclude this section, we find that the tQGT framework fully characterizes charge response in the transport limit for a general many-body state. This limit accounts for all responses in an insulator, but in phases with gapless charged excitations, there are additional contributions which will be the focus of the next section.

\section{Fermi surface contributions}
\label{sec:fs}
The goal of this section is to elucidate Fermi surface contributions to the response functions and tQGT.
We consider metals with well-defined quasiparticle bands, and evaluate all quantities with a Slater-determinant ground state built from the single-particle Green's function $G(k,i\omega)$, with the appropriate current and density operators that satisfy the continuity equation.

Let us begin with the simplest Fermi surface contribution that arises in the longitudinal conductivity.
The Drude weight, or charge stiffness, is given by the non-analytic part of the current-current bubble
\begin{equation}
    D_{\mu\nu} = \sum_m \intBZ \left( - \frac{\partial f}{\partial \varepsilon_{m}} \right) (\partial_{k_\mu} \varepsilon_{m,k})(\partial_{k_\nu} \varepsilon_{m,k}),
    \label{eq:drude:bubble}
\end{equation}
where $ m $ denotes the band index, $ k $ is the crystal momentum in the Brillouin zone (BZ), $\varepsilon_{m,k}$ is the dispersion and $ f $ is the Fermi-Dirac distribution.

We move next to the tQGT itself. At $T=0$, the many-body ground state is a Slater determinant of all Bloch states $\{ |\psi_{m,k}\rangle \}$ with energies below the Fermi energy $\{ \varepsilon_{m,k} < E_F \}$.
The dipole matrix elements between these Bloch states are given by
\begin{equation}
    \langle \psi_{m,k} | r_\mu | \psi_{n,k+q} \rangle = \delta(q) r_{\mu,k}^{mn} - i\delta_{mn} \partial_\mu \delta(q),
\end{equation}
where $r_{\mu,k}^{mn} \equiv \langle u_{m,k} | i \partial_\mu u_{n,k} \rangle$ is the dipole matrix element between the cell-periodic parts of the Bloch wavefunctions, $|\psi_{m,k}\rangle = e^{i k\cdot\hat{r}} | u_{m,k}\rangle$.
We then find that $\mathcal{Q}_{\mu\nu}(t)$ can be split into intra- and inter-band parts $\mathcal{Q}_{\mu\nu}(t) = \mathcal{Q}^{\rm intra}_{\mu\nu}(t) + \mathcal{Q}^{\rm inter}_{\mu\nu}(t)$
which are explicitly given by
\begin{equation}
    \mathcal{Q}^{\rm inter}_{\mu\nu}(t) = \intBZ \sum\limits_{m\neq n} f_{n,k} (1-f_{m, k} ) \;e^{ i \omega_{mn,k} t } r_{\mu,k}^{mn} r_{\nu,k}^{nm}
\end{equation}
and
\begin{align}
    &\mathcal{Q}^{\rm intra}_{\mu\nu}(t) = \sum\limits_{m} \intBZ dk dq \; f_{m,k}( 1 - f_{m,k+q}) \notag \\
    & \hspace{0.72cm} \times e^{ i( \varepsilon_{m,k} - \varepsilon_{m,k+q} )t } \left(i \partial_\mu \delta(q) \right) \langle \psi_{m,k+q} | \hat{r}_\nu | \psi_{m,k} \rangle.
\end{align}
Here we introduced $\omega_{mn,k} \equiv \varepsilon_{m,k} - \varepsilon_{n,k}$ to simplify the notation.
The inter-band part is well-behaved and has already been shown to capture the quantum geometry of insulators.
On the other hand, the intra-band part has divergent contributions from the Fermi surface. To that end, we shift the derivative from the delta function by using integration by parts and then use the identity
\begin{equation}
    \lim\limits_{q\rightarrow 0} \langle \psi_{m,k+q} | \hat{r}_\mu | \psi_{m,k} \rangle (\varepsilon_{m,k+q} - \varepsilon_{m,k}) = -i\,\partial_{k_\mu} \varepsilon_{m,k} \label{eq:iden-1}
\end{equation}
to break apart the intra-band contribution into two terms (Appendix~\ref{app:intra})
\begin{equation}
    \mathcal{Q}^{\rm intra}_{\mu\nu}(t) = D_{\mu\nu} t + \tilde{F}_{\mu\nu}.
\end{equation}
The linear-in-time component corresponds to the Drude weight of Eq.~\eqref{eq:drude:bubble}, consistent with the discussion in the previous section. However, it is not the only contribution from the Fermi surface. The second term, which is time-independent, is divergent
\begin{equation}
    \tilde{F}_{\mu\nu} = \sum\limits_m \intBZ \partial_{k_\mu} f_{m,k} \, \langle \psi_{m,k} | r_\nu | \psi_{m,k} \rangle.
\end{equation}
The divergence is evident in this expression. The dipole matrix elements are evaluated with Bloch states, which are plane waves. Since the Fermi factor restricts the momentum integration to the Fermi surface, $k \in {\rm FS}$, and the expectation value of position in a plane wave basis is ill-defined, this quantity is divergent.
Being time-independent, $\tilde F_{\mu\nu}$ drops from the conductivity, which involves only $\partial_t \mathcal{Q}^{\rm as}$. It does not drop from the equal-time tensor itself. It is the divergent metallic quantum weight, the same divergence found in the electronic localization length of the metallic state \cite{Resta1999, Souza2000}, and it carries no signature in dc transport. An equivalent form of $\tilde F_{\mu\nu}$, obtained without invoking Eq.~\eqref{eq:iden-1}, is derived in Appendix~\ref{app:intra}, and makes explicit that the divergence is a property of the Fermi surface alone.

We now return to the Hall response, whose Fermi surface content lies beyond the tQGT.
The transport limit of Hall conductivity is
\begin{equation}
    \lim\limits_{\omega \rightarrow 0} \epsilon_{\mu\nu} \dfrac{ {\rm Im}[\Lambda_{\mu\nu}(\omega)] }{\omega} = \sum\limits_{m} \intBZ f_{m,k} \Omega_{m,k}
\end{equation}
where $\Omega_{m,k} = 2\epsilon_{\mu\nu}{\rm Im}[ r^{mn}_{\mu,k}  r^{nm}_{\nu,k} ]$ is the Berry curvature, built from the dipole matrix elements introduced above. The other limit is relatively more difficult as it requires the $q_\mu$ derivative of the response function
\begin{equation}
    \chi_{0,\nu}(q) = \sum_{m,n} \intBZ \mathcal{F}^{mn}_{k,q} \langle u_{m,k+q} | u_{n,k} \rangle \langle u_{n,k} | j_\nu | u_{m,k+q} \rangle,
\end{equation}
where $\mathcal{F}^{mn}_{k,q} = (f_{m,k+q} - f_{n,k} )/( \varepsilon_{n,k} - \varepsilon_{m,k+q})$ is the Fermi factor that comes from the Matsubara sum. Unlike the transport limit, note that the sum includes the diagonal term $m = n$.

The derivative with respect to $ q_\mu $ can act on three components: $ \mathcal{F}^{mn}_{k,q} $, $ \langle u_{m,k+q} | u_{n,k} \rangle $, and $ \langle u_{n,k} | j_\nu | u_{m,k+q} \rangle $. When it acts on $ \mathcal{F}^{mn}_{k,q} $, the remaining terms simplify to $ \delta_{mn} $ and $ \partial_\mu \varepsilon_{n,k} $, which are symmetric in $ \mu $ and $ \nu $ and do not contribute.
The action of other two gives finite contributions. After a tedious but straightforward calculation (Appendix~\ref{app:hall}), we find that the Hall conductivity is given by
\begin{equation}
    \sigma_H = \intBZ \sum_m f_{m,k} \Omega_{m,k} + \left( -\frac{\partial f}{\partial \varepsilon_m} \right) \mu_{m,k} \label{eq:def:SigmaH:fs}
\end{equation}
with both Fermi sea and Fermi surface contributions. Note that $\mu_{m,k}$ is the orbital magnetic moment of band $m$ at momenta $k$
\begin{equation}
    \mu_{m,k} = \epsilon_{\mu\nu} \sum\limits_{n\neq m} {\rm Im}[\hat{r}_\mu^{mn} \hat{r}_\nu^{nm}] ( \varepsilon_{m,k} - \varepsilon_{n,k} ).
\end{equation}
Two observations follow. First, the Hall conductivity reduces to the TKNN formula in insulators, where the Fermi surface is absent. Second, the explicit Fermi surface term reveals the central role of the orbital magnetic moment in gapless systems, a contribution anticipated by Haldane in the context of the anomalous Hall effect of metals \cite{Haldane2004, Wang2007}; a complementary geometric route to the Drude weight and orbital magnetization, via quantum-geometric bounds, was recently given in Ref.~\cite{Shinada2025}.

\begin{figure*}
    \centering
    \includegraphics[width=2\columnwidth]{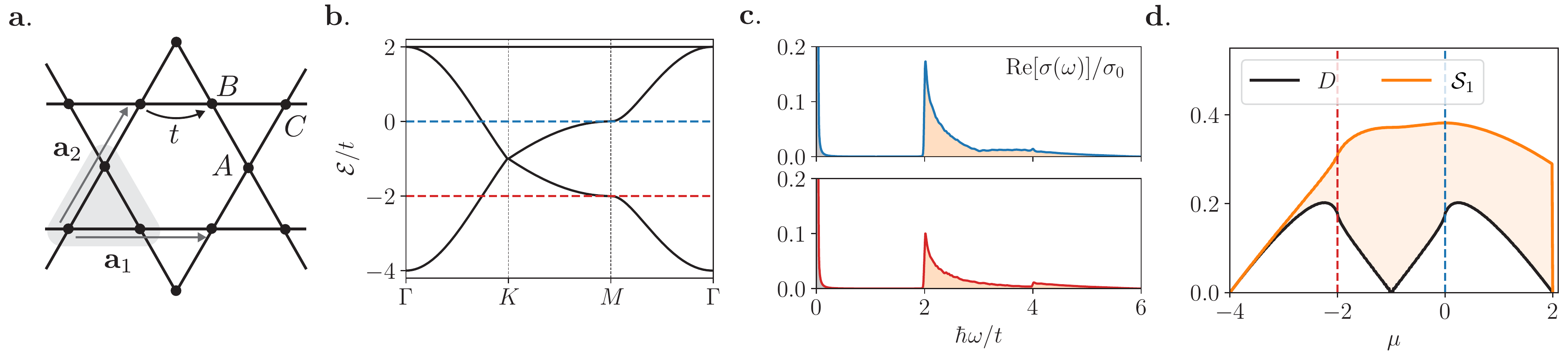}
    \caption{ Separation of scales in the kagome metal. (a) Kagome lattice with sublattices $A$, $B$, $C$, lattice vectors $\mathbf{a}_{1,2}$, and nearest-neighbor hopping $t$. (b) Band structure along $\Gamma$--$K$--$M$--$\Gamma$, with the flat band at $\varepsilon = 2t$. Dashed lines mark the two van Hove singularities, at $\mu = -2t$ (red) and $\mu = 0$ (blue), positioned symmetrically about the Dirac point. (c) Optical conductivity at the two van Hove fillings. The Fermi surfaces are identical and the Drude peaks nearly coincide, but the inter-band absorption (shaded) is markedly stronger at the upper van Hove filling, closer to the flat band. (d) Drude weight $D$ and total spectral weight $\mathcal{S}_1$ as a function of chemical potential $\mu$. The bound-charge reservoir $\mathcal{S}_1 - D$ (shaded) grows toward the flat band, with $D/\mathcal{S}_1 = 0.46$ at the van Hove filling adjacent to the flat band versus $0.57$ at its partner, despite identical Fermi surfaces.}
    \label{fig:kagome}

\end{figure*}

It can be argued that the Fermi surface contribution to the Hall conductivity is inconsequential for metals as it is not quantized.
However, this contribution is critical in getting the correct bulk magnetization.
Recall that the magnetization is defined as the change in Free energy with an applied magnetic field
\begin{equation}
    M = \dfrac{\partial F}{ \partial B} = \dfrac{\partial F}{ \partial n} \dfrac{ \partial n }{ \partial B}
\end{equation}
where $F$ is the Free energy of the system. In the second expression, we have introduced density and identified the first term as the change in chemical potential and the second term as the Hall conductivity. As a result, the total magnetization of the system is given by the integral
\begin{equation}
    M(E_F) = \int\limits_{-\infty}^{E_F} d\mu\; \sigma_H(\mu) = \intBZ \sum\limits_{m} f_{m,k} M_{m,k}
\end{equation}
where we have considered the distribution function $f_{m,k} = \Theta(\mu - \varepsilon_{m,k})$ and the band resolved components $M_{m,k}$ are
\begin{align}
    M_{m,k} &= \mu_{m,k} - \Omega_{m,k}(\varepsilon_{m,k} - E_F) \nonumber \\
    &= \epsilon_{\mu\nu} \sum\limits_{n \neq m} {\rm Im}[ \hat{r}_\mu^{nm} \hat{r}_\nu^{mn} ] \,(2E_F - \varepsilon_{m,k} - \varepsilon_{n,k}),
\end{align}
in agreement with the modern theory of orbital magnetization \cite{Xiao2005, Thonhauser2005, Shi2007}. Although not quantized on its own, the magnetization varies linearly with $E_F$ as it is varied inside the gap of a topological insulator, with the slope given by the Chern number, $\Delta M/\Delta E_F \propto \mathcal{C}$.
We emphasize that this result could not have been obtained without the Fermi surface contribution that follows from Eq.~\eqref{eq:sigmaH:metal}
\begin{equation}
    \sigma_H^M = \intBZ \sum_m \left( -\frac{\partial f}{\partial \varepsilon_m} \right) \mu_{m,k}.
\end{equation}

\section{Itinerant and bound electrons}
\label{sec:itinerant}
The preceding sections have laid the foundation of the tQGT framework and its connection to Fermi surface contributions. In this section, we will focus on the transport limit and the longitudinal conductivity, where tQGT captures all contributions, and use it as a tool to understand the dynamics of electrons.

Identifying a suitable low-energy basis for an interacting system is highly nontrivial. Ab initio methods such as DFT provide single-particle states, but constructing an effective low-energy theory requires a minimal tight-binding description. In many cases, due to factors ranging from topology to geometric frustration, the minimal model has multiple orbitals.
Once the tight-binding basis is established, the projectors $\hat{P}$ and $\hat{Q} = 1- \hat{P}$ are both naturally defined, providing the framework for tQGT.

We begin our discussion with the $f$-sum rule, which follows from the time derivative of the tQGT
\begin{equation}
    \mathcal{S}_1 = \int\limits_0^\infty d\omega \;  {\rm Re}[\sigma(\omega)] = \dfrac{\pi e^2}{\hbar} \left[ (-i \hat{\partial}_t) \mathcal{Q}(t) \right]_{t=0}.
\end{equation}
The total spectral weight $\mathcal{S}_1$ consists of two contributions. The first arises from electrons near the Fermi surface and is associated with the charge stiffness $D$. The second contribution originates from bound electrons, which we denote as $D_g = \mathcal{S}_1 - D$ and refer to as the bound-charge weight. We note that $\mathcal{S}_1$ is the energy-weighted moment of the conductivity; the quantum metric itself enters through the inverse-energy-weighted moment ${\rm Re}[\mathcal{Q}(0)]$, which diverges in a metal. $D_g$ is its finite, sum-rule-protected counterpart, and measures the inter-band dipole weight available at the lattice scale \cite{Basov2005}.

The separation between itinerant and bound electrons can be seen explicitly in the optical conductivity, illustrated in Fig.~\ref{fig:mainIdea}, where intra-band and inter-band transitions contribute to distinct absorption peaks. While the precise separation depends on the scattering time $\tau$, the spectral weights $D$ and $\mathcal{S}_1$ do not depend on $\tau$.
We can then use the ratio $D/\mathcal{S}_1$ to infer the proportion of electrons that are mobile to those that are bound.
The latter are a reservoir of electrons that are immobile, but can become itinerant when perturbed.

As an example, we consider the kagome lattice, shown in Fig.~\ref{fig:kagome}. With three sublattices per unit cell, it hosts three bands in momentum space with varying degrees of hybridization. The dispersive bands resemble those of graphene and appear to exhibit particle-hole symmetry about the Dirac point. However, this symmetry does not extend to the wavefunctions.

To illustrate this, we examine the two van Hove singularities symmetrically positioned around the Dirac point. Despite their identical Fermi surfaces, the ratio of itinerant to bound electrons $D/\mathcal{S}_1$ differs by roughly twenty percent: for the nearest-neighbor model we find $D/\mathcal{S}_1 = 0.46$ at the van Hove filling adjacent to the flat band ($E_F = 0$) and $D/\mathcal{S}_1 = 0.57$ at its partner ($E_F = -2t$).
The asymmetry is not a statement about energetics or joint density of states. The sum rule counts all transitions irrespective of frequency, so proximity to the flat band in energy cannot by itself change $\mathcal{S}_1$. The asymmetry lives in the wavefunctions \cite{Kiesel2012}. The saddle point adjacent to the flat band inherits the sublattice-interference structure responsible for the flat band, and with it stronger inter-band dipole matrix elements.
This distinction is reflected in calculations as well, where doping to the van Hove near the flat band leads to correlated phases, while doping the van Hove further from the flat band results in a weakly correlated system \cite{Kiesel2013, Wang2013kagome, Liu2024lowervH}.

This motif is universal across materials, all of which host deep-lying atomic states that contribute to optical conductivity.
The central feature of our formalism is that the transitions responsible for $D/\mathcal{S}_1$ originate from orbital hybridization at the lattice scale, and do not include atomic transitions, which occur at much higher energies.
Tight-binding models are of course an approximation that retains only a select set of orbitals of a real material, which discretizes the position operator and restricts the dipole transitions that appear in optical conductivity. Including all orbitals is impractical, and also unnecessary, as many are filled and inert to the low-energy physics.
The central idea behind our discussion of the sum rule $\mathcal{S}_1$ is to identify low-energy spectral weight that can flow to zero frequency, as observed most strikingly in flat-band lattices \cite{Peotta2015, Torma2021, Checkelsky2024}.
While our formalism does not specify the mechanism underlying this transfer, it identifies the necessary condition for it to occur, namely the presence of a bound-charge reservoir at the lattice scale. Several microscopic mechanisms have already been proposed and are the subject of active investigation \cite{Ahn2021transfer, Carmichael2025, Antebi2024flatband, Burkov2026QuantumGeometricDiffusion}.

\section{Topological semimetals}
Within the tQGT framework, we identify two defining characteristics of a metal: a finite charge stiffness, which gives rise to a linear-in-time component, and a divergent time-independent contribution. Both stem from the presence of a Fermi surface. Semimetals require more care. At the undoped node both Fermi-surface quantities vanish, $D = 0$ and the intra-band $\tilde F = 0$, since there are no states at the Fermi level; the singular behavior resides instead in the \emph{inter-band} part of the tQGT, which inherits the nodal structure of the dipole matrix elements. Doping restores a small Fermi surface, and with it a finite $D$ and a divergent $\tilde F$.
Since the conductivity is non-analytic in semimetals, we use the definition where zero frequency limit is taken before zero temperature.

For a doped semimetal with a small but finite Fermi surface, we can rewrite the intra-band components of tQGT after performing the Fermi-surface integrals
\begin{equation}
    D = N(0) \langle v^2 \rangle_{\rm FS}, \quad \tilde{F} = N(0) \langle v\cdot \hat{r} \rangle_{\rm FS}
\end{equation}
where $\hat{r}$ is the position operator, $v$ is the group velocity and $\langle \cdot \rangle_{\rm FS}$ denotes an average over the Fermi surface. Clearly, the group velocity is well-behaved and hence the charge stiffness vanishes as the density of states at the Fermi level approaches zero, $N(0) \rightarrow 0$. On the other hand, the position operator is highly singular. The product $N(0)\langle v \cdot \hat{r} \rangle_{\rm FS}$ is therefore a competition in which the divergence of the position matrix element wins for any finite Fermi surface, however small, while at the undoped node the intra-band contribution vanishes identically.
The precise form of divergence depends on the dimensions of the system and the details of the band structure.

The inter-band part of tQGT has special properties as well. The quantum geometry of nodal semimetals has been studied extensively in the frequency domain \cite{Ahn2020}, and general frameworks for time-dependent quantum geometry have recently been formulated \cite{Guan2026}. Closed-form time-domain results for gapless nodal systems have, to our knowledge, not been reported; we provide them here.
To illustrate, we consider a Dirac system in 2D described by the Hamiltonian
\begin{equation}
    H_k = \hbar v_F k \cdot \sigma
\end{equation}
where $v_F$ is the Dirac velocity and $\sigma$ is a vector of Pauli matrices in the sublattice space. The system has two bands with
\begin{equation}
    \varepsilon_{\pm,k} =\pm \hbar v_F |k|, \quad  |u_{\pm, k}\rangle = \dfrac{1}{\sqrt{2}} \begin{pmatrix}
        1 \\ \pm e^{ i \varphi_k}
    \end{pmatrix}
\end{equation}
where $\varphi_k$ is the polar angle of $k$. These result in the dipole matrix element
\begin{equation}
    |\langle u_{+,k} | i \partial_k u_{-,k} \rangle|^2 = \dfrac{1}{4 |k|^2} \label{eq:dirac:dipMat}
\end{equation}
and the inter-band tQGT
\begin{align}
    \mathcal{Q}^{\rm inter}(t) &= \int\limits_0^\infty \dfrac{dk}{2\pi} \dfrac{ e^{ 2i\hbar v_F k t } }{ 4 k } \\
    &= \dfrac{-1}{8\pi}\left( \gamma + \ln 2 \hbar v_F |t| - \dfrac{i\pi}{2} {\rm sgn}(t)\right)
\end{align}
where $\gamma$ is the Euler constant. The integral is logarithmically divergent at small $k$; a lower cutoff, set by finite doping or system size, supplies the scale inside the logarithm, and all cutoff-dependent constants drop from $\partial_t \mathcal{Q}^{\rm as}$, leaving the physical response cutoff independent. This expression proves that a Dirac theory has dipole fluctuations at all scales, and is consistent with the minimal conductivity of Dirac fermions \cite{Ludwig1994, Ando2002, Katsnelson2006, Nair2008}. The anti-symmetric part of tQGT only couples to the sign function ${\rm sgn}(t)$ and yields
\begin{equation}
    \sigma(t) = \dfrac{\pi e^2}{\hbar} \Theta(t) \partial_t \mathcal{Q}^{\rm as}(t) = \dfrac{\pi e^2}{4\hbar} \delta(t)
\end{equation}
which in Fourier space gives a frequency independent contribution.
As a result, we conclude that because of scale invariance in dipole fluctuations there is no separation of scales.
We note that doping a Dirac system introduces Fermi energy $E_F$ into the system which naturally leads to a separation of scales.

While similar conclusions extend to other semimetals, the precise balance between these quantities is not guaranteed. Two-dimensional Dirac fermions are the marginal case, where the logarithmic tQGT recovers the known frequency-independent conductivity of graphene; the scale invariance follows from linear dispersion in two dimensions and is destroyed by any infrared scale, such as doping or a gap.
As a concrete example, we consider a 3D Weyl semimetal, which has the same dipole matrix elements as those in Eq.~\eqref{eq:dirac:dipMat}.
However, due to the increased phase-space measure, the resulting expression for the interband tQGT is
\begin{equation}
    \mathcal{Q}^{\rm inter}(t) = \int\limits_0^\infty \dfrac{dk}{8\pi^2} e^{ +2i\hbar v_F k t } = \dfrac{i}{16 \pi^2 \hbar v_F t}
\end{equation}
which shows that dipole fluctuations are suppressed at long times.
While there is a formal divergence at short times, this singularity is an artifact of the unbounded nature of the spectrum and disappears upon introducing a physical cutoff or regularization scheme. Importantly, the universal aspect of a semi-metal is its asymptotic behavior at long times.

\section{Discussion}
The tQGT provides a correlator of projected position operators from which all transport-limit charge responses of a many-body state follow. In this work we extended it to gapless phases, where it acquires two Fermi surface features. The first is a linear-in-time component equal to the charge stiffness, and second is a divergent time-independent term that cancels in every physical response. The stiffness measures dipole fluctuations that spread linearly in time, with $D$ playing the role of Drude weight; the divergence reflects the delocalized nature of the metallic ground state and is the same divergence found in the electronic localization length of metals \cite{Resta1999, Souza2000}. Both statements are properties of the order of limits in the current response and hold beyond single-particle theory.

Responses in which the limits $q \rightarrow 0$ and $\omega \rightarrow 0$ do not commute lie beyond the tQGT. In the longitudinal channel gauge invariance closes this gap through the $f$-sum rule, but in the Hall channel a genuinely metallic correction $\sigma_H^{\rm M}$ survives, given by the orbital magnetic moment of states at the Fermi surface. This term, anticipated by Haldane \cite{Haldane2004}, distinguishes the Streda response of a metal from its transport Hall conductivity and is required to recover the correct bulk orbital magnetization \cite{Xiao2005}. Establishing Eq.~\eqref{eq:sigmaH:metal} at the many-body level, beyond the perturbative argument given here, remains an open problem.

For materials, one output of this framework is the ratio $D/\mathcal{S}_1$ of itinerant to total low-energy spectral weight. This ratio is fixed by orbital hybridization at the lattice scale rather than by energetic proximity between bands. In the kagome lattice, the two van Hove singularities have identical Fermi surfaces yet sharply different $D/\mathcal{S}_1$. The bound-charge reservoir quantified by $\mathcal{S}_1 - D$ is spectral weight that can be transferred to low frequencies, and we propose $D/\mathcal{S}_1$ as a simple probe for metals in which such geometry-assisted spectral weight transfer is possible. Topological semimetals occupy the opposite extreme. In two-dimensional Dirac systems dipole fluctuations are scale invariant and no separation between bound and itinerant charge exists, while in Weyl semimetals the fluctuations decay as $1/t$ and the separation is restored at long times.

\paragraph*{Acknowledgements.---}
This work is supported by the Alfred P. Sloan Foundation (FG-2025-24714), and the NSF CAREER program (DMR-2340394).
The Flatiron Institute is a division of the Simons Foundation.

\appendix

\section{Intra-band tQGT and the divergent term}
\label{app:intra}
We derive the decomposition $\mathcal{Q}^{\rm intra}_{\mu\nu}(t) = D_{\mu\nu} t + \tilde{F}_{\mu\nu}$. Dropping the band label, the anti-symmetrized intra-band tQGT reads
\begin{widetext}
\begin{equation}
    \mathcal{Q}^{\rm intra, as}_{\mu\nu}(t) = \dfrac{1}{2i} \int dk \int dq \; \partial_\mu \delta(q) \Big( f_{k}(1 - f_{k+q} )  e^{ i( \varepsilon_{k} - \varepsilon_{k+q} )t }  \langle \psi_{k+q} | r_\nu | \psi_{k} \rangle - f_{k+q}(1 - f_{k} )  e^{ i( \varepsilon_{k+q} - \varepsilon_{k} )t }  \langle \psi_{k} | r_\nu | \psi_{k+q} \rangle \Big).
\end{equation}
\end{widetext}
Integrating by parts in $q$ and letting the derivative act on the exponential, the occupations, and the matrix element generates one time-dependent and two time-independent pieces. For the time-dependent piece, the identity in Eq.~\eqref{eq:iden-1} converts the singular position matrix element into the group velocity, and the $q \rightarrow 0$ limit gives
\begin{equation}
    \sum_m \intBZ \left( - \frac{\partial f}{\partial \varepsilon_{m}} \right) (\partial_{k_\mu} \varepsilon_{m,k})(\partial_{k_\nu} \varepsilon_{m,k}) \, t = D_{\mu\nu}\, t,
\end{equation}
recovering the charge stiffness as the coefficient of the linear-in-time growth of dipole fluctuations; the factor $-i$ in Eq.~\eqref{eq:iden-1}, combined with the $-i$ in the dipole decomposition, renders the coefficient of $t$ real. The time-independent pieces combine into
\begin{equation}
    \tilde{F}_{\mu\nu} = \sum\limits_m \intBZ \partial_{k_\mu} f_{m,k} \, \langle \psi_{m,k} | r_\nu | \psi_{m,k} \rangle,
\end{equation}
quoted in the main text, which is divergent because the expectation value of the position operator in a Bloch state is ill-defined.

An equivalent form is obtained by acting with the $q$ derivative directly on the occupation factors and the states, without invoking Eq.~\eqref{eq:iden-1}. The occupation term gives
\begin{equation}
    \dfrac{1}{2} \int dk \; (-\partial_\mu f_k) \langle \psi_{k} | r_\nu | \psi_{k} \rangle,
\end{equation}
while the term where the derivative acts on the state reduces, at finite temperature, to
\begin{equation}
    i \int dk \; f_k(1-f_{k} ) \, {\rm Im}[\langle \partial_\mu \psi_{k} | r_\nu | \psi_k \rangle].
\end{equation}
The factor $f_k(1-f_k)$ is supported on the thermal shell around the Fermi surface and vanishes pointwise as $T \rightarrow 0$; the two forms of $\tilde F_{\mu\nu}$ agree only when the zero-temperature limit is taken after the regularization of the position matrix element, which is the order used throughout. This makes explicit that the divergence of $\tilde{F}_{\mu\nu}$ is a Fermi surface property: it is absent in any gapped state and independent of the interior of the Fermi sea.

\section{Fermi surface contributions to the Hall conductivity}
\label{app:hall}
Here we derive Eq.~\eqref{eq:def:SigmaH:fs} by evaluating the $q_\mu$ derivative of the density-current response function
\begin{equation}
    \chi_{0,\nu}(q) = \sum_{m,n} \intBZ \mathcal{F}^{mn}_{k,q} \langle u_{m,k+q} | u_{n,k} \rangle \langle u_{n,k} | j_\nu | u_{m,k+q} \rangle.
\end{equation}
As discussed in the main text, the derivative acting on $\mathcal{F}^{mn}_{k,q}$ produces a term symmetric in $\mu \leftrightarrow \nu$ that drops out upon anti-symmetrization. The remaining contributions come from the overlap and from the current matrix element.

Acting on the overlap $\langle u_{m,k+q} | u_{n,k} \rangle$ yields
\begin{equation}
    \sum_{m,n} \intBZ \mathcal{F}^{mn}_{k} \langle \partial_\mu u_{m,k} | u_{n,k} \rangle \langle u_{n,k} | j_\nu | u_{m,k} \rangle,
\end{equation}
which contains both a Fermi surface piece from the diagonal $m=n$ term, where $\mathcal{F}^{mm}_k$ approached $-\partial f/\partial\varepsilon$ and $\langle u_{m,k}|j_\nu|u_{m,k}\rangle = \partial_\nu \varepsilon_{m,k}$,
\begin{equation}
   [\sigma_H]_1 = i\epsilon_{\mu\nu} \sum_{m} \intBZ \left(- \dfrac{\partial f}{\partial \varepsilon} \right) \langle \partial_\mu u_{m,k} | u_{m,k} \rangle \, \partial_\nu \varepsilon_{m,k},
\end{equation}
and an inter-band piece
\begin{equation}
    [\sigma_H]_2 = i\epsilon_{\mu\nu} \sum_{m\neq n} \intBZ \mathcal{F}^{mn}_{k} \langle \partial_\mu u_{m,k} | u_{n,k} \rangle \langle u_{n,k} | j_\nu | u_{m,k} \rangle.
\end{equation}
Combining the velocity with the derivative of the Fermi function, $(-\partial f/\partial\varepsilon)\partial_\nu\varepsilon_{m,k} = -\partial_\nu f_{m,k}$, and identifying the Berry connection $\mathcal{A}_{\mu,m,k} = i\langle u_{m,k} | \partial_\mu u_{m,k} \rangle$, the Fermi surface piece becomes, after integration by parts,
\begin{equation}
    [\sigma_H]_1 = \epsilon_{\mu\nu} \sum_{m} \intBZ f_{m,k} \, \partial_\mu \mathcal{A}_{\nu, m, k}.
\end{equation}
This is the Fermi-surface Berry-phase contribution to the anomalous Hall effect identified by Haldane \cite{Haldane2004}. For the inter-band piece, we use $\langle u_{n,k} | j_\nu | u_{m,k} \rangle = (\varepsilon_{m,k} - \varepsilon_{n,k}) \langle u_{n,k} | \partial_\nu u_{m,k} \rangle$ for $m \neq n$ to cancel the energy denominator in $\mathcal{F}^{mn}_k$, giving
\begin{equation}
    [\sigma_H]_2 = \sum_{m} \intBZ f_{m,k} \, \Omega_{m,k},
\end{equation}
Note that $[\sigma_H]_1$ does not survive to the final answer: it is cancelled exactly by the first term of $[\sigma_H]_3$ below, so the Fermi-sea curvature term is counted once, through $[\sigma_H]_2$.

Finally, the derivative acting on the current matrix element gives
\begin{equation}
    [\sigma_H]_3 = i\epsilon_{\mu\nu} \sum_{m} \intBZ \left(- \dfrac{\partial f}{\partial \varepsilon} \right) \langle u_{m,k} | j_\nu | \partial_\mu u_{m,k} \rangle,
\end{equation}
whose integrand decomposes as
\begin{align}
     &\langle u_{m,k} | j_\nu | \partial_\mu u_{m,k} \rangle = \partial_\nu \varepsilon_{m,k} \langle u_{m,k} | \partial_\mu u_{m,k} \rangle \notag \\
    &+ \sum\limits_{n\neq m} \langle u_{m,k} | \partial_\nu u_{n,k} \rangle( \varepsilon_{n,k} - \varepsilon_{m,k} ) \langle u_{n,k} | \partial_\mu u_{m,k} \rangle.
\end{align}
The first term cancels $[\sigma_H]_1$ exactly. The second defines the orbital magnetic moment of band $m$,
\begin{equation}
    \mu_{m,k} = \epsilon_{\mu\nu} \sum\limits_{n\neq m} {\rm Im}[\hat{r}_\mu^{mn} \hat{r}_\nu^{nm}] ( \varepsilon_{m,k} - \varepsilon_{n,k} ),
\end{equation}
which carries no factor of $2$, in contrast to the Berry curvature $\Omega_{m,k} = 2\epsilon_{\mu\nu} {\rm Im}[r^{mn}_{\mu,k} r^{nm}_{\nu,k}]$. Adding the three pieces gives Eq.~\eqref{eq:def:SigmaH:fs},
\begin{align}
    \sigma_H &= [\sigma_H]_1 + [\sigma_H]_2 + [\sigma_H]_3 \nonumber \\
    &= \intBZ \sum_m f_{m,k} \Omega_{m,k} + \left( -\frac{\partial f}{\partial \varepsilon} \right) \mu_{m,k}.
\end{align}

\bibliography{tqgtmetals}

\end{document}